\documentstyle[12pt,epsf]{article}
\newcommand{\del}{\partial}

\newcommand{\half}{{1 \over 2}}
\newcommand{\st}{{\tilde t}}
\newcommand{\sq}{{\tilde q}}
\newcommand{\absv}[1]{\left|#1\right|}
\newcommand{\Dmt}{{m_{\tilde q}^2}}
\newcommand{\cc}{{{3g_2^2-g_1^2}\over{12}}}
\newcommand{\gtsim}{\mathrel{\hbox{\raise0.2ex
\hbox{$>$}\kern-0.75em\raise-0.9ex\hbox{$\sim$}}}}
\newcommand{\ltsim}{\mathrel{\hbox{\raise0.2ex
\hbox{$<$}\kern-0.75em\raise-0.9ex\hbox{$\sim$}}}}
\newcommand{\lw}[1]{\smash{\lower2.0ex\hbox{#1}}}
\newcommand{\PRD}[3]{Phys. Rev. {\bf D{#1}} (19{#2}) {#3}}
\newcommand{\PRLet}[3]{Phys. Rev. Lett. {\bf {#1}} (19{#2}) {#3}}
\newcommand{\NPB}[3]{Nucl. Phys. {\bf B{#1}} (19{#2}) {#3}}
\newcommand{\PLB}[3]{Phys. Lett. {\bf B{#1}} (19{#2}) {#3}}
\newcommand{\PrTP}[3]{Prog. Theor. Phys. {\bf {#1}} (19{#2}) {#3}}
\makeatletter

\@addtoreset{equation}{section}
\makeatother
\begin{document}
\begin{flushright}
SAGA--HE--134\\
March 25,~1998
\end{flushright}
\vspace{40pt}
\centerline{\Large{\bf Spontaneous $CP$ Violation at Finite Temperature}}
\centerline{\Large{\bf in the MSSM}\footnote{Talk by K.~Funakubo 
at the Workshop on Fermion Mass and $CP$ Violation, Hiroshima, Japan,
March 1998.}}
\vskip1.2cm
\begin{center}
{\bf Koichi~Funakubo$^{a,}$\footnote{e-mail: funakubo@cc.saga-u.ac.jp},
 Akira~Kakuto$^{b,}$\footnote{e-mail: kakuto@fuk.kindai.ac.jp},\\
 Shoichiro~Otsuki$^{b,}$\footnote{e-mail: otks1scp@mbox.nc.kyushu-u.ac.jp}
 and Fumihiko~Toyoda$^{b,}$\footnote{e-mail: ftoyoda@fuk.kindai.ac.jp}}
\end{center}
\begin{center}
{\it $^{a)}$Department of Physics, Saga University,
Saga 8408502 Japan}
\vskip 0.2 cm
{\it $^{b)}$Kyushu School of Engineering, Kinki University,
Iizuka 8208255 Japan}
\end{center}
\vskip 0.5 cm
\begin{quote}{\small
By studying the effective potential of the MSSM at finite temperature,
we find that $CP$ can be spontaneously broken in the intermediate 
region between the symmetric and broken phases separated by the bubble
wall created at the phase transition. 
If $CP$ is violated in this manner, there could be a bubble wall 
connecting $CP$ conserving vacua and violating $CP$ halfway, which
would result in sufficient baryon asymmetry of the universe.
}
\end{quote}
\section{Introduction}
Within the framework of electroweak theory, the baryon asymmetry of 
the universe (BAU) could be explained if the electroweak phase 
transition (EWPT) in the early universe is of first order\cite{reviewEB}.
The minimal standard model with sufficiently heavy Higgs boson 
($m_H>67\mbox{GeV}$) is found to have second-order EWPT.
Further the effect of the only $CP$ violation in the KM matrix is
too weak to generate the present BAU. Hence, one is led to extend the
standard model in such a way that the extra source of $CP$ violation
exists and the bosonic content is enlarged to have the EWPT stronger.
Among such extensions, the minimal supersymmetric standard model 
(MSSM) with the small soft-supersymmetry-breaking parameter in 
the stop mass-squared matrix\cite{MSSM-PT}\  will cause the first-order
EWPT and offer many sources of $CP$ violation.
The $CP$ violations effective for generation of the chiral charge flux
at the tree level are the relative phases among the $\mu$-parameter,
the gaugino mass parameters and the $A$-parameters included in the
mass matrices of the charginos, neutralinos and the scalar partners
of the quarks and leptons\cite{EWB-MSSM}.\ 
Besides these phases, the relative phase of the two Higgs doublets
not only enters these mass matrices but also directly couple to
the quarks and leptons to yield the flux carried by these particles
into the symmetric phase region\cite{NKC,FKOTTa}.
In general, the phases of the Higgs scalars vary spatially around
the bubble wall so that the gauge-invariant relative phase cannot be
rotated away from the Yukawa interactions.\par
In a previous paper\cite{FKOTTb}\  we attempted to determine the profile 
of the bubble wall by solving the equations of motion for the effective 
potential at the transition temperature ($>T_C\simeq100\mbox{GeV}$) in
the two-Higgs-doublet model.
For some set of parameters, we presented a solution such that $CP$-violating 
phase spontaneously generated becomes as large as $O(1)$ 
around the wall while it completely vanishes in the broken and 
symmetric phase limits. We shall refer to this mechanism as
`transitional $CP$ violation'.
This solution gives a significant hypercharge flux, by the quark or 
lepton transport\cite{FKOTTc}.\  
We also showed that a tiny explicit $CP$ violation, which is consistent with 
the present bound on the neutron EDM, does nonperturbatively resolve the
degeneracy between the $CP$-conjugate pair of the bubbles to leave a sufficient
BAU after the EWPT\cite{FKOTa}.\par
We examine the possibility of the transitional $CP$ 
violation at finite temperature in the MSSM by studying the effective
potential, when one of the stop mass parameters is small so that
the high-temperature expansion is applied only to it.
\section{Transitional $CP$ violation}
Consider a model with two Higgs doublets whose VEVs are parameterized 
as
\begin{equation}
  \langle\Phi_i\rangle
 =\left(\begin{array}{c} 0 \\
                        {1\over{\sqrt2}}\rho_i e^{i\theta_i}
        \end{array}\right),\qquad(i=1,2)
\end{equation}
and $\theta\equiv\theta_1-\theta_2$.
We assume that the gauge-invariant effective potential near the 
transition temperature has the form of
\begin{eqnarray}
 V_{\rm eff}(\rho_i,\theta_i)&=&
 \half m_1^2\rho_1^2+\half m_2^2\rho_2^2 - m_3^2\rho_1\rho_2\cos\theta
 +{{\lambda_1}\over8}\rho_1^4+{{\lambda_2}\over8}\rho_2^4   \nonumber  \\
&+&{{\lambda_3+\lambda_4}\over4}\rho_1^2\rho_2^2
 + {{\lambda_5}\over4}\rho_1^2\rho_2^2\cos2\theta
 - \half(\lambda_6\rho_1^2+\lambda_7\rho_2^2)\rho_1\rho_2\cos\theta
    \nonumber  \\
&-& [A\rho_1^3+\rho_1^2\rho_2(B_0+B_1\cos\theta+B_2\cos2\theta)  
    \nonumber  \\
& &\qquad
    + \rho_1\rho_2^2(C_0+C_1\cos\theta+C_2\cos2\theta) + D\rho_2^3 ],
    \nonumber  \\
&=&\left[{\lambda_5\over2}\rho_1^2\rho_2^2
         -2(B_2\rho_1^2\rho_2+C_2\rho_1\rho_2^2)\right]  \nonumber\\
& &\qquad\times
   \left[\cos\theta-{{2m_3^2+\lambda_6\rho_1^2+\lambda_7\rho_2^2+
  2(B_1\rho_1+C_1\rho_2)}\over{2\lambda_5\rho_1\rho_2-8(B_2\rho_1+C_2\rho_2)}}
   \right]^2     \nonumber  \\
& & +\theta\mbox{-independent terms}.  \label{eq:Veff-ansatz}
\end{eqnarray}
The $\rho^3$-terms are expected to be induced at finite temperature 
in a model whose EWPT is of first order.
Since we do not consider any explicit $CP$ violation, all the 
parameters are assumed to be real.
For a given $(\rho_1,\rho_2)$, the spontaneous $CP$ violation occurs if
\begin{eqnarray}
  F(\rho_1,\rho_2) &\equiv&
  \displaystyle{
  {\lambda_5\over2}\rho_1^2\rho_2^2-2(B_2\rho_1^2\rho_2+C_2\rho_1\rho_2^2)
  }>0,       \label{eq:def-F}\\
 -1 <G(\rho_1,\rho_2) &\equiv&
 \displaystyle{
 {{2m_3^2+\lambda_6\rho_1^2+\lambda_7\rho_2^2+2(B_1\rho_1+C_1\rho_2)}\over
  {2\lambda_5\rho_1\rho_2-8(B_2\rho_1+C_2\rho_2)}}
 } <1.       \label{eq:def-G}
\end{eqnarray}
In the MSSM at the tree level, $\lambda_5=\lambda_6=\lambda_7=0$ and
$A=B_k=C_k=D=0$ ($k=0,1,2$), so that no spontaneous $CP$ violation occurs.
At zero temperature ($A=B_k=C_k=D=0$), it is argued that $\lambda_{5,6,7}$
are induced radiatively and (\ref{eq:def-F}) is satisfied if the 
contributions from the chargino and neutralino are large 
enough. For (\ref{eq:def-G}) to be satisfied, $m_3^2$ should be as small as
$\lambda_6\rho_1^2+\lambda_7\rho_2^2$ so that the pseudoscalar becomes 
too light\cite{Maekawa},\par
At $T\simeq T_C$, the values of $(\rho_1,\rho_2)$ vary
from $(0,0)$ to $(v_C\sin\beta_C,v_C\cos\beta_C)$ between the symmetric and 
broken phase regions, where the subscript $C$ denotes the quantities 
at the transition temperature. Then the effective parameters in 
(\ref{eq:Veff-ansatz}) include the temperature corrections as 
well. Hence there arises large possibility to satisfy both 
(\ref{eq:def-F}) and (\ref{eq:def-G}) in the intermediate region at
the transition temperature, without accompanying too light scalar.
If this is the case, a local minimum or a valley of $V_{\rm eff}$ 
appears for intermediate $(\rho_1,\rho_2)$ with a nontrivial 
$\theta$.
For such a $V_{\rm eff}$ with appropriate effective 
parameters, the equations of motion for the Higgs fields predict that  
some class of solutions exist, which have $\theta$ of $O(1)$ in the 
intermediate region even if it vanishes in the broken phase\cite{FKOTTa}.\par
\section{Effective parameters of the MSSM}
Since we are concerned with the possibility of 
the spontaneous $CP$ violation, all the parameters in the lagrangian
are assumed to be real.
The tree-level Higgs potential of the MSSM is
\begin{eqnarray}
 V_0 &=&
 m_1^2\varphi_d^\dagger\varphi_d + m_2^2\varphi_u^\dagger\varphi_u
 +(m_3^2\varphi_u\varphi_d + \mbox{h.c})     \nonumber\\
 & &
 +{{\lambda_1}\over2}(\varphi_d^\dagger\varphi_d)^2
 +{{\lambda_2}\over2}(\varphi_u^\dagger\varphi_u)^2
 +\lambda_3(\varphi_u^\dagger\varphi_u)(\varphi_d^\dagger\varphi_d)
 +\lambda_4(\varphi_u\varphi_d)(\varphi_u\varphi_d)^* \nonumber\\
 & &
 +\left[
  {{\lambda_5}\over2}(\varphi_u\varphi_d)^2 +
  (\lambda_6\varphi_d^\dagger\varphi_d+\lambda_7\varphi_u^\dagger\varphi_u)
  \varphi_u\varphi_d + \mbox{h.c}
  \right],
\end{eqnarray}
where
\begin{eqnarray}
 m_1^2 &=& \tilde{m}_d^2+\absv{\mu}^2,\qquad
 m_1^2  =  \tilde{m}_u^2+\absv{\mu}^2,\qquad
 m_3^2  =  \mu B,   \nonumber\\
 \lambda_1&=&\lambda_2={1\over4}(g_2^2+g_1^2),\qquad
 \lambda_3 = {1\over4}(g_2^2-g_1^2),\qquad
 \lambda_4 = -\half g_2^2,    \\
 \lambda_5 &=& \lambda_6 = \lambda_7 = 0,
\end{eqnarray}
Here $g_{2(1)}$ is the $SU(2)$($U(1)$) gauge coupling, $\mu$ is the
coefficient of the Higgs quadratic interaction in the superpotential.
The mass squared parameters $\tilde{m}_{u,d}^2$ and $\mu B$ come from the
soft-supersymmetry-breaking terms so that they are arbitrary at this
level. $m_3^2$ could be complex but its phase can be eliminated by the
redefinition of the fields when $\lambda_5=\lambda_6=\lambda_7=0$.
We adopt the convention in which this $m_3^2$ is real and positive.\par
Let us parameterize the VEVs of the Higgs doublets as 
\begin{equation}
  \varphi_d
 ={1\over{\sqrt2}}\pmatrix{\rho_1\cr 0}
 ={1\over{\sqrt2}}\pmatrix{ v_1\cr 0},\qquad
  \varphi_u
 ={1\over{\sqrt2}}\pmatrix{ 0\cr \rho_2 e^{i\theta}}
 ={1\over{\sqrt2}}\pmatrix{ 0\cr v_2+iv_3}.  \label{eq:VEV}
\end{equation}
The effective potential at the one-loop level is
\begin{equation}
 V_{\rm eff}=V_0+V_1(\rho_i,\theta)+{\bar V}_1(\rho_i,\theta;T),
\end{equation}
where $V_1(\rho_i,\theta)$ is the zero-temperature correction given by
\begin{equation}
 V_1(\rho_i,\theta) =
 \sum_j\,n_j {{m_j^4}\over{64\pi^2}}
 \left[\log\left({{m_j^2}\over{M_{\rm ren}^2}}\right)-{3\over2}
 \right],       \label{eq:V1-zero-T}
\end{equation}
and ${\bar V}_1(\rho_i,\theta;T)$ is the finite temperature 
correction;
\begin{equation}
 {\bar V}_1(\rho_i,\theta;T) =
 {T^4\over2\pi^2}\sum_j\, n_j\int_0^\infty dx\,x^2
 \log\left[ 1 -{\rm sgn}(n_j)
  \exp\left(-\sqrt{x^2+m_j^2/T^2}\right)\right].
                \label{eq:V1-finite-T}
\end{equation}
Here we used the $\overline{\rm DR}$-scheme to renormalize $V_{\rm eff}$
with the renormalization scale $M_{\rm ren}$.
$n_j$ counts the degrees of freedom of each species including its 
statistics, that is, $n_j>0$ ($n_j<0$) for bosons (fermions).
$m_j$, which is a function of the Higgs background $(\rho_i,\theta)$,
is the mass eigenvalue of each species.\par
At the one-loop level, $\left(m_3^2\right)_{\rm eff}$ receives 
corrections only from the Higgs bosons, squarks, sleptons, and 
charginos and neutralinos. $\lambda_{5,6,7}$, which are zero at
the tree level, are generated only through the loops of these particles.
Among them, we consider the contributions of charginos($\chi^{\pm}$), 
neutralinos($\chi^0$), stops($\st$) and Higgs($\phi^{\pm}$).
The effective parameters are defined as the derivatives of $V_{\rm eff}$
at the origin of the order-parameter space:
\begin{eqnarray}
 \left(m_3^2\right)_{\rm eff} &=&
 -\left.{{\del^2 V_{\rm eff}}\over{\del v_1\del v_2}}\right|_0 =
 m_3^2 + \Delta_{\chi}m_3^2+ \Delta_{\tilde t}m_3^2 + 
               \Delta_{\phi^\pm}m_3^2,    \label{eq:def-eff-m32}\\
 \lambda_5 &=&
 \half\left(
  \left.{{\del^4 V_{\rm eff}}\over{\del v_1^2\del v_2^2}}\right|_0 -
  \left.{{\del^4 V_{\rm eff}}\over{\del v_1^2\del v_3^2}}\right|_0
 \right) =
 \Delta_{\chi}\lambda_5 + \Delta_{\tilde t}\lambda_5 +
               \Delta_{\phi^\pm}\lambda_5, \label{eq:def-eff-l5}\\ 
 \lambda_6 &=&
 -{1\over3}
   \left.{{\del^4 V_{\rm eff}}\over{\del v_1^3\del v_2}}\right|_0 =
 \Delta_{\chi}\lambda_6 + \Delta_{\tilde t}\lambda_6 +
               \Delta_{\phi^\pm}\lambda_6, \label{eq:def-eff-l6}\\ 
 \lambda_7 &=&
 -{1\over3}
   \left.{{\del^4 V_{\rm eff}}\over{\del v_1\del v_2^3}}\right|_0 = 
 \Delta_{\chi}\lambda_7 + \Delta_{\tilde t}\lambda_7 +
               \Delta_{\phi^\pm}\lambda_7. \label{eq:def-eff-l7}
\end{eqnarray}
Here we show only the expressions of the contributions from the 
charginos, neutralinos and stops, since that from the charged Higgs
is much smaller than the others for our choice of the parameters in
the next section. As for the derivation, see \cite{FKOTc}.\par
When the gaugino mass parameters satisfy $M_2=M_1$, the contributions
from the charginos have the same factors as those from the neutralinos.
Then
\begin{eqnarray}
 \Delta_{\chi}m_3^2 &=&
 -2g_2^2\left(1+{1\over{\cos^2\theta_W}}\right)\mu M_2\, L(M_2,\mu) 
 +{{g_2^2}\over{\pi^2}}\left(1+{1\over{\cos^2\theta_W}}\right)\mu M_2
  f_2^{(+)}\!\left({{M_2}\over T},{\mu\over T}\right), \nonumber\\
 & &              \label{eq:m32-gaugino}\\
 \Delta_{\chi}\lambda_5 &=&
  {{g_2^4}\over{8\pi^2}}\left(1+{2\over{\cos^4\theta_W}}\right)
     K\!\left({{M_2^2}\over{\mu^2}}\right)
 -{{g_2^4}\over{\pi^2T^4}}\left(1+{2\over{\cos^4\theta_W}}\right)\mu^2 M_2^2
     f_4^{(+)}\!\left({{M_2}\over T},{\mu\over T}\right), \nonumber\\
 & &              \label{eq:l5-gaugino}\\
 \Delta_{\chi}\lambda_6 &=&
 -{{g_2^4}\over{8\pi^2}}\left(1+{2\over{\cos^4\theta_W}}\right)
  {\mu\over{M_2}}\left[ - H\!\left({{M_2^2}\over{\mu^2}}\right)
      + K\!\left({{M_2^2}\over{\mu^2}}\right) \right]   \nonumber\\
 & &\qquad +
 {{g_2^4}\over{\pi^2}}\left(1+{2\over{\cos^4\theta_W}}\right) \left[
  {{\mu M_2}\over{T^2}}
          f_3^{(+)}\!\left({{M_2}\over T},{\mu\over T}\right)
  +{{\mu^3 M_2}\over{T^4}}
           f_4^{(+)}\!\left({{M_2}\over T},{\mu\over T}\right)\right]
                                  \nonumber\\
 &=& \Delta_{\chi}\lambda_7,        \label{eq:l6-gaugino}
\end{eqnarray}
where functions $f^{(\pm)}_n(a,b)$, $L(m_1,m_2)$, $K(\alpha)$ and 
$H(\alpha)$ are defined in \cite{FKOTc}. For the condition $F>0$,
it is essential to have a positive $\lambda_5$. The dominant positive 
contribution comes from (\ref{eq:l5-gaugino}), so that
$\mu^2\simeq M_2^2$ is required since $K(\alpha)$ is maximum at
$\alpha=1$.\par
The stop contributions are divided into two parts, one of which comes
from the light stop, whose supersymmetry-breaking mass parameters are 
small, while the other originates from the heavier mass eigenstates.
\begin{eqnarray}
 \Delta_{\st}m_3^2 &=&
  N_c y_t^2\mu A_t\, L(m_\sq,0) +
  {{3T^2}\over{\pi^2}}{{y_t^2\mu A_t}\over\Dmt}\left[
   I_B^\prime(a_\sq^2)-{{\pi^2}\over12}\right],     \label{eq:m32-stop2}\\
 \Delta_{\st}\lambda_5 &=&
 -{{N_cy_t^4}\over{16\pi^2}}{{\mu^2A_t^2}\over{m_\sq^2M_{IR}^2}}
     K\left({{m_\sq^2}\over{M_{IR}^2}}\right)  \nonumber\\
 & &{}+
 {{N_cy_t^4\mu^2A_t^2}\over{\pi^2(\Dmt)^2}}\left[
  {{2T^2}\over\Dmt}\left( 
     -I_B'(a_\sq^2)+{\pi^2\over12} \right) + I_B''(a_\sq^2) + 2\lambda_- 
     \right],                 \label{eq:l5-stop2}\\
 \Delta_{\st}\lambda_6 &=&
  {{N_cy_t^2}\over{16\pi^2}}{{\mu A_t}\over{m_\sq^2}}\left[
  {1\over4}\left({{g_1^2}\over3}-g_2^2\right) -
  {{g_1^2m_\sq^2}\over{3M_{IR}^2}}H\left({{M_{IR}^2}\over{m_\sq^2}}\right)+
  {{y_t^2\mu^2}\over{M_{IR}^2}}
     K\left({{m_\sq^2}\over{M_{IR}^2}}\right) \right] \nonumber\\
 & &+
 {{N_cy_t^2\mu A_t}\over{\pi^2\Dmt}}\left\{
   {{2T^2}\over\Dmt}\left({{y_t^2\mu^2}\over\Dmt}+(-{5\over3}g_1^2+g_2^2)
      \right)\left[I_B'(a_\sq^2)-{\pi^2\over12}\right]\right.\nonumber\\
 & &\qquad
 \left. -\left({{y_t^2\mu^2}\over\Dmt}+\cc\right)I_B''(a_\sq^2)
     +2\left({{g_1^2}\over3}-{{y_t^2\mu^2}\over\Dmt}
       \right)\lambda_- \right\},          \label{eq:l6-stop2}\\
 \Delta_{\st}\lambda_7 &=&
  {{N_cy_t^2}\over{16\pi^2}}{{\mu A_t}\over{m_\sq^2}}\Biggl[
  -\left(y_t^2+{1\over4}\left({{g_1^2}\over3}-g_2^2\right)\right)
   \nonumber\\
  & &{}-\left(y_t^2-{{g_1^2}\over3}\right){{m_\sq^2}\over{M_{IR}^2}}
        H\left({{M_{IR}^2}\over{m_\sq^2}}\right)
  +{{y_t^2 A_t^2}\over{M_{IR}^2}}
      K\left({{m_\sq^2}\over{M_{IR}^2}}\right) \Biggr] \nonumber\\
  & &{}+
 {{N_cy_t^2\mu A_t}\over{\pi^2\Dmt}}\Biggl\{
   {{2T^2}\over\Dmt}\left({{y_t^2\mu^2}\over\Dmt}-(-{5\over3}g_1^2+g_2^2)
 \right)\!\!
     \left[I_B'(a_\sq^2)-{\pi^2\over12}\right] \nonumber\\
 & &\qquad
  -\left(y_t^2+{{y_t^2\mu^2}\over\Dmt}-\cc\right)\!
        I_B''(a_\sq^2)
   +2\left(y_t^2-{{g_1^2}\over3}-{{y_t^2\mu^2}\over\Dmt}\right)\lambda_-
     \Biggr\},          \label{eq:l7-stop2}
\end{eqnarray}
where $N_c=3$ and $M_{IR}$ is the infrared cutoff parameter, 
which will be taken to be the order of the transition temperature.
$I_B(a^2)$ is defined in \cite{FKOTc}.
\section{Numerical Results}
We examine whether the conditions $G(\rho_1,\rho_2)<1$ and 
$\absv{G(\rho_1,\rho_2)}<1$ are satisfied or not by evaluating the
effective parameters included in $F$ and $G$. 
We take $v_0=246\mbox{GeV}$, $m_t=177\mbox{GeV}$, 
$M_{\rm ren}=\mbox{GeV}$ and $\tan\beta_0=5$. The mass parameters
in the Higgs potential is given by the relations which determine
the tree-level minimum of the potential:
\begin{eqnarray}
 m_1^2 &=& m_3^2\tan\beta_0-{1\over2}m_Z^2\cos(2\beta_0), \nonumber\\
 m_2^2 &=& m_3^2\cot\beta_0+{1\over2}m_Z^2\cos(2\beta_0).
\end{eqnarray}
\par
\begin{figure}
 \epsfxsize=11.6cm
 \centerline{\epsfbox{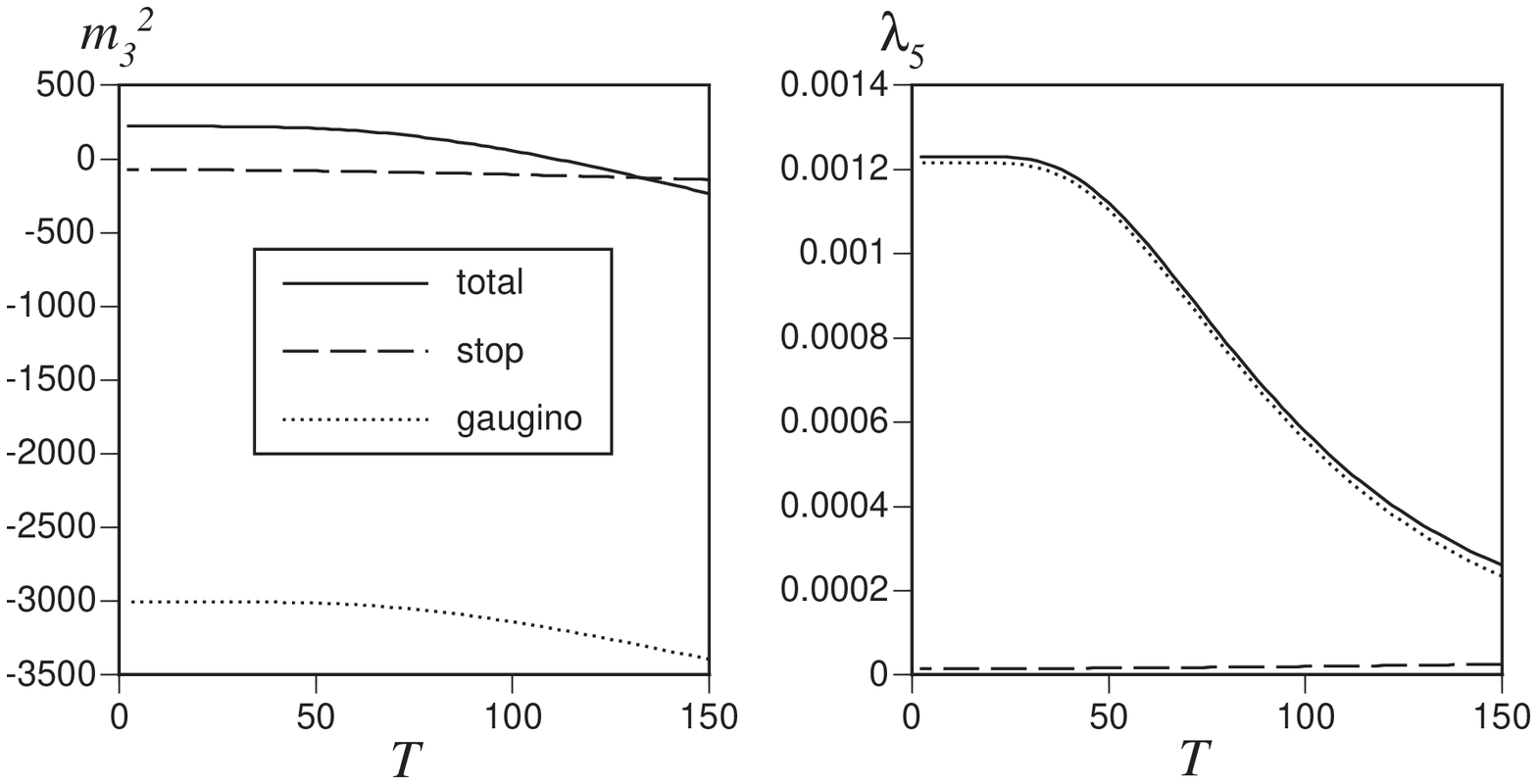}}
 \caption{$\left(m_3^2\right)_{\rm eff}$ and $\lambda_5$ as functions
 of temperature $T$. The total values are given by the solid curves, the 
 corrections from the stop, chargino-neutralino are depicted by 
 the dashed, dotted and dotted-dashed curves, respectively.
 }
 \label{fig:1}
\end{figure}

\begin{figure}
 \epsfxsize=11.6cm
 \centerline{\epsfbox{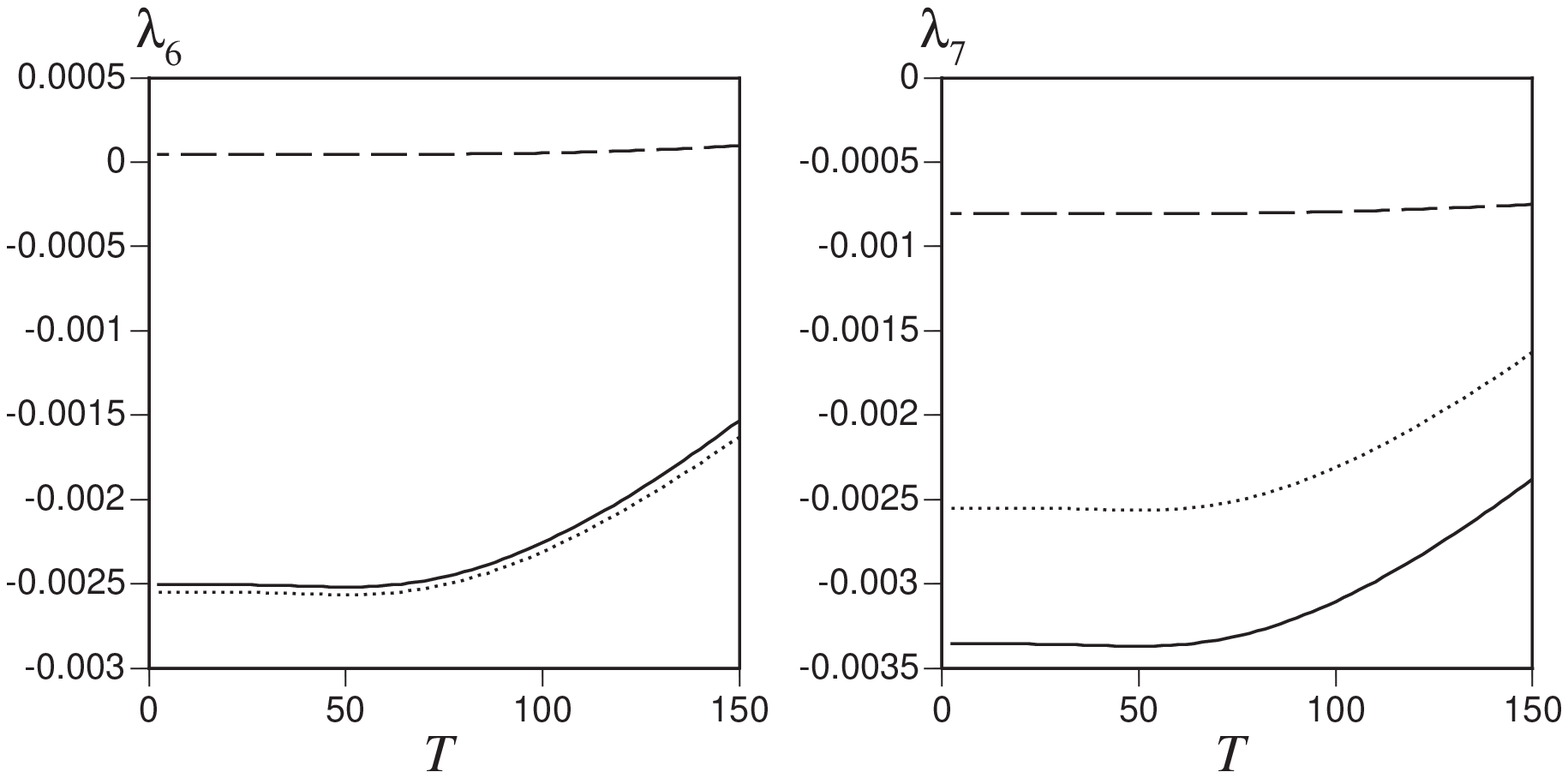}}
 \caption{$\lambda_6$ and $\lambda_7$ as functions
 of temperature.
 }
 \label{fig:2}
\end{figure}

\noindent
(A) $\mu A_t>0$\\
In this case, we take, as the remaining parameters, the values
in Table~\ref{table:1}.
The behaviors of the effective parameters are shown in the 
Figs.~\ref{fig:1} and \ref{fig:2}.

\begin{figure}
 \epsfxsize=11.6cm
 \centerline{\epsfbox{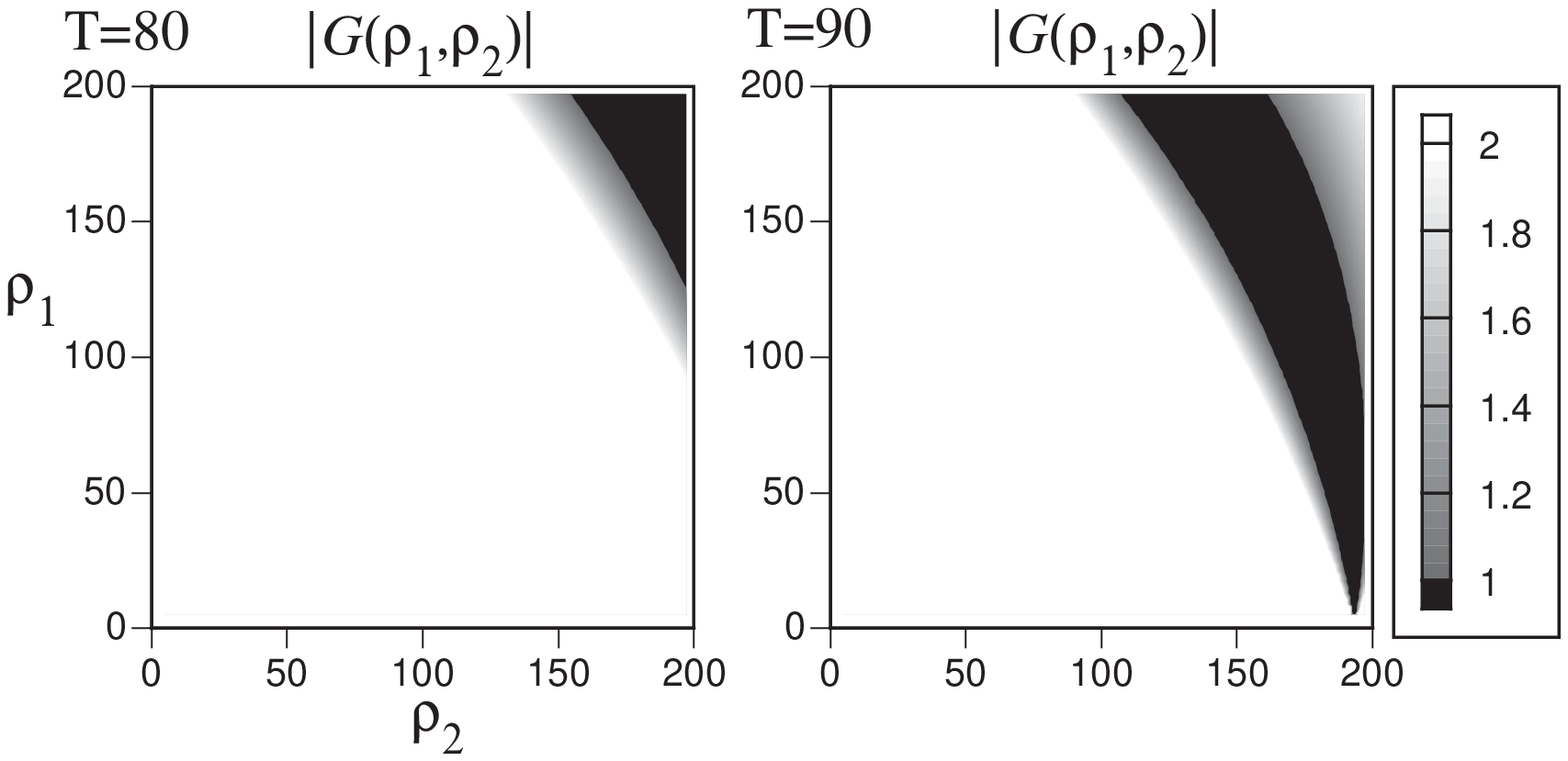}}
 \caption{Contour plots of $\absv{G(\rho_1,\rho_2)}$ at $T=80$ and 
 $90$. $\absv{G(\rho_1,\rho_2)}$ is satisfied in the black region.
 }
 \label{fig:3}
\end{figure}

\begin{center}
\begin{table}
 \caption{The parameters used in the numerical analysis in the case
 of $\mu A_t>0$.}
 \label{table:1}
 \begin{center}
  \begin{tabular}{ccccc}
   \hline
    $m_3^2$&$A_t$& $M_2$ & $\mu$ &$m_\sq$ \\
   \hline
    $3300$ & $10$& $-400$ &$200$ &  $400$ \\
   \hline
  \end{tabular}
 \end{center}
\end{table}
\end{center}
The behaviors of the effective parameters are shown in the 
Figs.~\ref{fig:1} and \ref{fig:2}.
In this case, we have
\begin{equation}
 B_2/T = 1.532346\times10^{-5},\qquad
 C_1/T = -4.846485\times10^{-3},
\end{equation}
so that $F(\rho_1,\rho_2)>0$ is satisfied for $\rho_2>5.104$ at 
$T=100\mbox{GeV}$. The region in which $\absv{G(\rho_1,\rho_2)}<1$ is 
satisfied are depicted in Figs.~\ref{fig:3} and \ref{fig:3}.
\begin{figure}
 \epsfxsize=10cm
 \centerline{\epsfbox{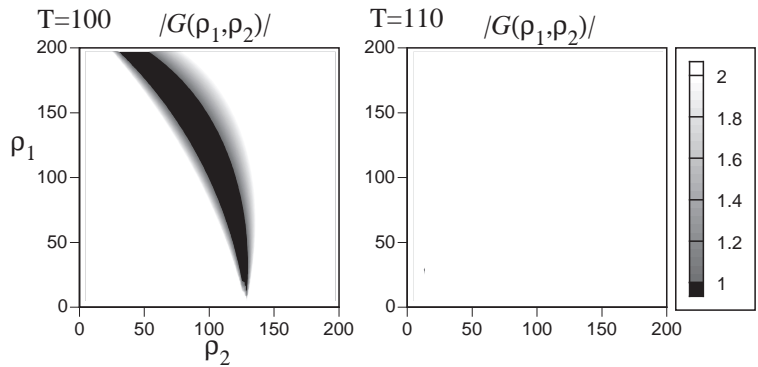}}
 \caption{The same as Fig.~\protect\ref{fig:3} at $T=100$ and $110$. 
 }
 \label{fig:4}
\end{figure}
\begin{table}
 \caption{The parameters used in the numerical analysis in the case
 of $\mu A_t<0$.}
 \label{table:2}
 \begin{center}
  \begin{tabular}{ccccc}
   \hline
    $m_3^2$&$A_t$& $M_2$ & $\mu$ &$m_\sq$ \\
   \hline
    $2200$ & $10$& $300$ &$-300$ &  $400$ \\
   \hline
  \end{tabular}
 \end{center}
\end{table}
\par
\begin{figure}
 \epsfxsize=10cm
 \centerline{\epsfbox{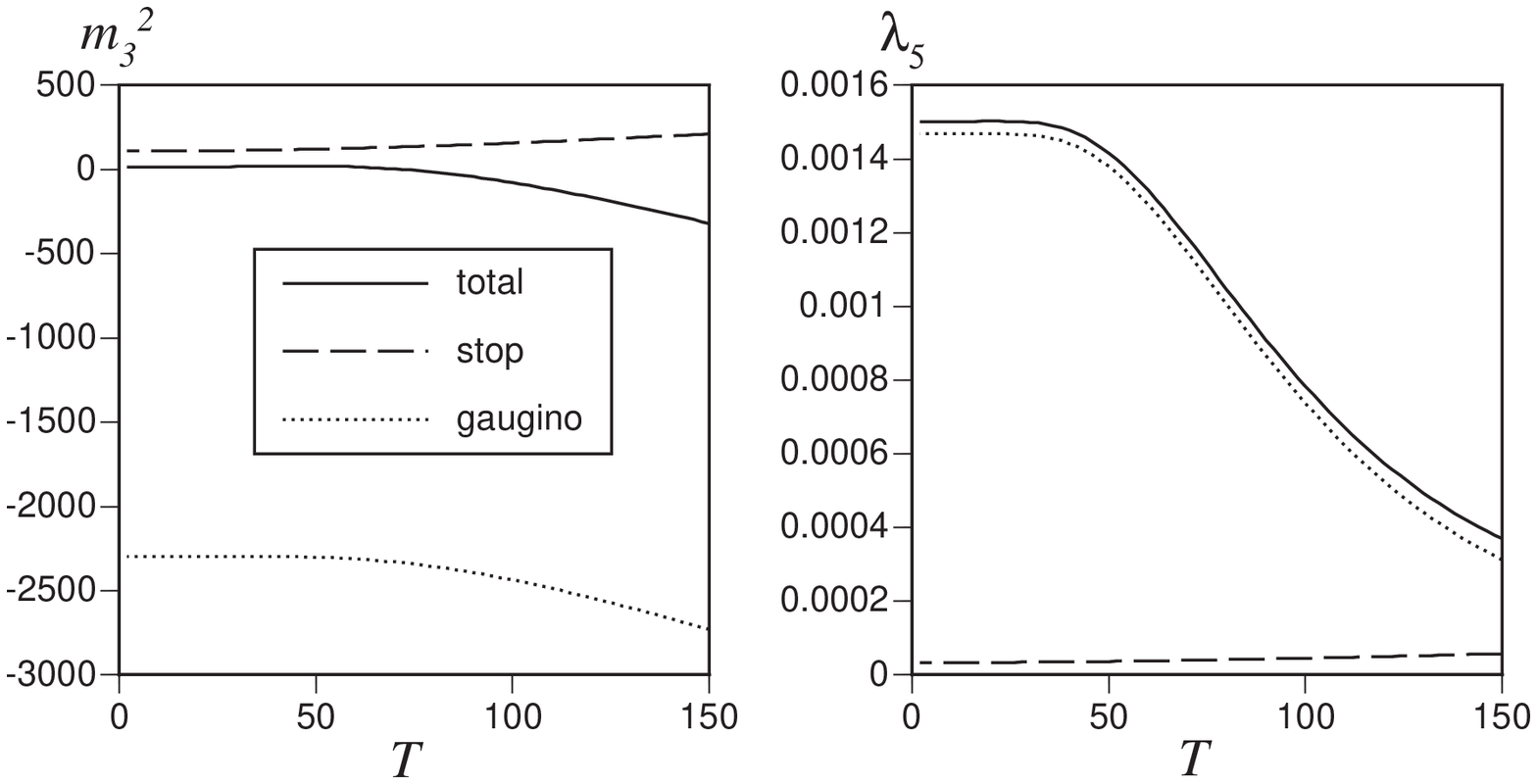}}
 \caption{$\left(m_3^2\right)_{\rm eff}$ and $\lambda_5$ as functions
 of temperature. The total values are given by the solid curves, the 
 corrections from the stop and chargino-neutralino are depicted by
 the dashed and dotted curves, respectively.
 }
 \label{fig:5}
\end{figure}

\begin{figure}
 \epsfxsize=10cm
 \centerline{\epsfbox{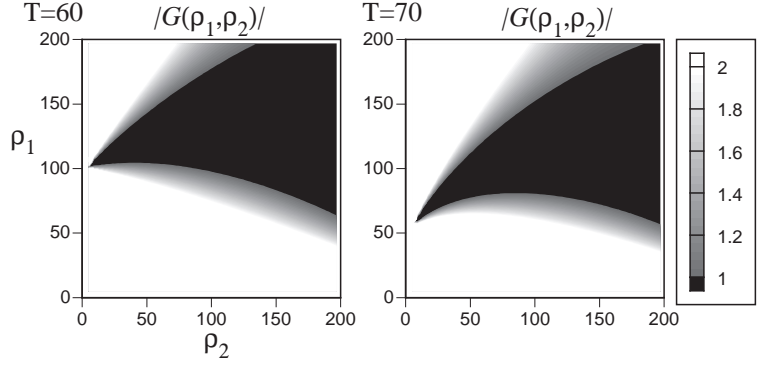}}
 \caption{Contour plots of $\absv{G(\rho_1,\rho_2)}$ at $T=60$ and 
 $70$. $\absv{G(\rho_1,\rho_2)}$ is satisfied in the black region.
 }
 \label{fig:7}
\end{figure}

\begin{figure}
 \epsfxsize=10cm
 \centerline{\epsfbox{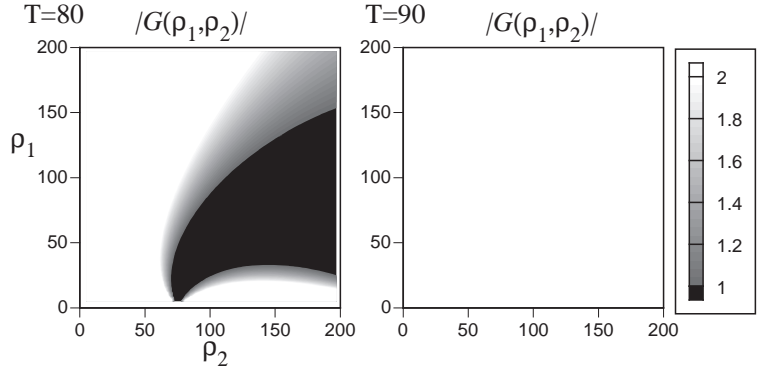}}
 \caption{The same as Fig.~\protect\ref{fig:7} at $T=80$ and $90$. 
 }
 \label{fig:8}
\end{figure}

\noindent
(B) $\mu A_t<0$\\
The input parameters are given in Table~\ref{table:2}.
The temperature-dependences of the effective parameters are shown
in Figs.~\ref{fig:5} and \ref{fig:5}.
\begin{figure}
 \epsfxsize=10cm
 \centerline{\epsfbox{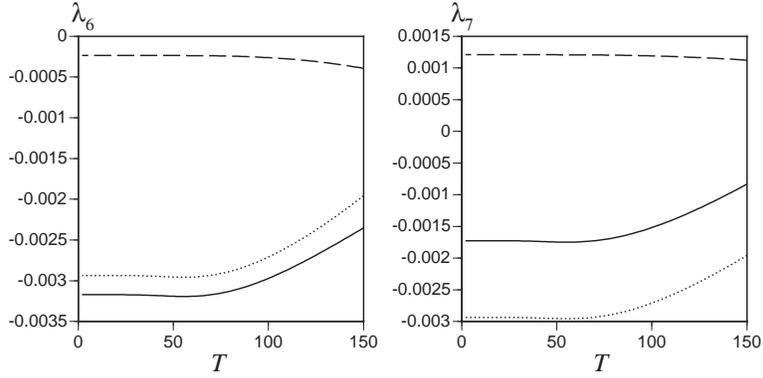}}
 \caption{$\lambda_6$ and $\lambda_7$ as functions
 of temperature.
 }
 \label{fig:6}
\end{figure}
For this parameter set, we have
\begin{equation}
 B_2/T = 1.657862\times10^{-5},\qquad
 C_1/T = 3.470457\times10^{-3},
\end{equation}
which implies $F(\rho_1,\rho_2)>0$ for $\rho_2>8.463$ at $T=100$.
The region in which $\absv{G(\rho_1,\rho_2)}<1$ is 
satisfied are depicted in Figs.~\ref{fig:7} and \ref{fig:8}.
Note that both $\lambda_6$ and $\lambda_7$ are negative, while
$C_1$ is negative for the case (A) and positive for (B).
This makes the $\rho$-dependence of $G(\rho_1,\rho_2)$ in the case (B) 
weaker than in the case (A). Hence The condition
$\absv{G(\rho_1,\rho_2)}<1$ is satisfied in broader region for the 
case (B).
\section{Discussions}
The two conditions for the transitional $CP$ violation is satisfied 
at $T\simeq T_C$, if (i) $\absv{\mu}\simeq\absv{M_{2,1}}$ for 
$\Delta_\chi\lambda_5>0$,  (ii) $\mu M_2<0$ and $m_3^2$ to decrease
$\left(m_3^2\right)_{\rm eff}$ by the $\chi$-contributions and
(iii) $m_3^2$ is tuned at the tree-level for $\left(m_3^2\right)_{\rm eff}$
to become as small as $\lambda_{6,7}\rho^2$.
This mechanism of $CP$ violation is attractive, since it will be free 
from the constraints on $CP$ violation at $T=0$, it can generate 
sufficient $BAU$ and it is not bothered by the light-scalar problem.
Note that for the parameters that admit the transitional $CP$ 
violation, $CP$ is {\it not} spontaneously broken at $T=0$.\par
It should be noted that the spontaneously-$CP$-breaking minimum
does not have to be the global minimum of $V_{\rm eff}$.
The transitional $CP$ violation could take place if the conditions
are satisfied for some fixed $(\rho_1,\rho_2)$, since such a bubble 
wall with transitional $CP$ violation would have a lower energy than
that without $CP$ violation.
For another reason, however, we should understand the global structure
of the effective potential, which determines $T_C$, to know whether
the transitional $CP$ violation occurs or not.
In this sense, the conditions we examined here should be regarded
as the necessary conditions but not the sufficient ones.
With the knowledge of the global structure of $V_{\rm eff}$, one could
find the $CP$-violating profile of the bubble wall so that one could
estimate the generated baryon number.
%
%

%
%

\baselineskip=13pt
\begin{thebibliography}{99}
\bibitem{reviewEB} For a review see, A.~Cohen, D.~Kaplan and A.~Nelson,
Ann. Rev. Nucl. Part. Sci. {\bf 43} (1993) 27.  \\
K.~Funakubo,\PrTP{96}{96}{475}.
\bibitem{MSSM-PT} M.~Carena, M.~Quiros, C.E.M.~Wagner, \PLB{380}{96}{81}.\\
D.~Delepine, J.-M.~Gerard, R.~Gonzalez Felipe, J.~Weyers,
\PLB{386}{96}{183}.\\
J.R.~Espinosa, \NPB{475}{96}{273}.\\
B.~de~Carlos and J.R.~Espinosa,\NPB{503}{97}{24}.
\bibitem{EWB-MSSM} P.~Huet and A.~Nelson, \PRD{53}{96}{4578}.\\
M.~Aoki, N~.Oshimo and A.~Sugamoto, \PrTP{98}{97}{1179}.\\
M.~P.~Worah, \PRLet{79}{97}{3810}.
\bibitem{NKC}  A.~Nelson, D.~Kaplan and A.~Cohen, \NPB{373}{92}{453}.
\bibitem{FKOTTa} K.~Funakubo, A.~Kakuto, S.~Otsuki, K.~Takenaga and F.~Toyoda,
\PRD{50}{94}{1105}.
\bibitem{FKOTTb} K.~Funakubo, A.~Kakuto, S.~Otsuki, K.~Takenaga and F.~Toyoda,
\PrTP{94}{95}{845}.
\bibitem{FKOTTc} K.~Funakubo, A.~Kakuto, S.~Otsuki, K.~Takenaga and F.~Toyoda,
\PrTP{93}{95}{1067}.
\bibitem{FKOTa} K.~Funakubo, A.~Kakuto, S.~Otsuki and F.~Toyoda,
\PrTP{96}{96}{771}.
\bibitem{FKOTb} K.~Funakubo, A.~Kakuto, ~S.~Otsuki, F.~Toyoda, 
\PrTP{98}{97}{427}.
\bibitem{Maekawa} N.~Maekawa, \PLB{282}{92}{387};
Nucl. Phys. Suppl. {\bf 37A} (1994) 191.
\bibitem{FKOTc} K.~Funakubo, A.~Kakuto, ~S.~Otsuki, F.~Toyoda,
hep-ph/9802276.
\end{thebibliography}
\end{document}